\documentclass[twoside,11pt]{article}
\usepackage{a4wide}
\usepackage[tbtags]{amsmath}
\usepackage{accents}
\usepackage{rotating}

\newlength{\lrvecw}

\numberwithin{equation}{section}

\newcommand{\beq}{\begin{equation}}
\newcommand{\eeq}{\end{equation}}
\newcommand{\beqa}{\begin{eqnarray}}
\newcommand{\eeqa}{\end{eqnarray}}
\newcommand{\beqar}{\begin{eqnarray*}}
\newcommand{\eeqar}{\end{eqnarray*}}

\newcommand{\calA}{{\cal A}}
\newcommand{\starM}{*_{\scriptscriptstyle M}}
\newcommand{\starN}{*_{\scriptscriptstyle N}}

\newcommand{\abar}{\bar{a}}
\def\msr{{\rm I\!R}} 

\begin{document}
\makeatletter
\title{Star Products and Perturbative Quantum Field Theory}
\author{Allen C. Hirshfeld\footnote{hirsh@physik.uni-dortmund.de}\, and 
Peter Henselder\footnote{henselde@dilbert.physik.uni-dortmund.de}\\
Fachbereich Physik\\
Universit\"at Dortmund\\
44221 Dortmund}

\maketitle


%

\begin{abstract}
{\small\sl  We discuss the deformation quantization approach to  perturbative quantum field theory. The various forms of Wick's theorem are shown to be a direct consequence of the structure of the star products. The scattering function for a scalar field in interaction with a spacetime-dependent source is derived. The translation of the results to the operator formalism reproduces known relations which lead to the derivation of the Feynman rules.}
\end{abstract}



\section{Introduction}
Deformation quantization as a framework for quantum physics has many attractive features. After setting up a classical system as a commutative algebra of functions on phase space, the passage to the corresponding quantum stystem is completely staightforward: the quantum observables are the same phase space functions as the classical observables. The only thing that changes is the product which is used to multiply phase space functions: in the quantum case one uses a {\em star product} instead of the commutative pointwise product of classical mechanics. The use of the star product produces all the characteristic quantum effects: non-commutativity, Heisenberg uncertainty and nonlocality. For a recent review at an elementary level see Ref. \cite{HH}.

When deformation quantization was first put forward as an autonomous program for quantum physics by Bayen et al. in 1978 \cite{Bayen} a few applications to non-relativistic quantum mechanics were known, but little concerning possible uses in relativistic quantum field theory. This situation has recently changed. After Dito's work on star products and free scalar fields \cite{D1}, D\"{u}tsch and Fredenhagen have studied the relation between deformation quantization and the algebraic approach to perturbative quantum field theory \cite{Fred}. Deformation theory has been used to establish important results concerning the quantization of gauge field theories, including the mathematical reconstruction of the BRST procedure \cite{Bordemann,Fred2} and the Batalin-Vilkoviski formalism \cite{Stasheff}. Cattaneo and Felder \cite{Cattaneo} have demonstrated a direct relationship between Kontsevich's star product on a general Poisson manifold \cite{Kontsevich} and the perturbative expansion of the Poisson-sigma model, which describes a two-dimensional topological field theory \cite{Strobl,Hirshfeld}.

The purpose of the present paper is threefold. Firstly, we wish to demonstrate that the deformation quantization approach, besides its mathematically appealing features, provides a {\em practical} method for dealing with problems of quantum physics. Secondly, we attempt to illuminate certain basic aspects of perturbative quantum field theory, by offering new insights into the structures which underlie the calculations. Finally, although our work certainly does not render the recent more specialized applications referred to above any the less remarkable, it may make them appear less surprising.

In the present paper we give a systematic account of the passage from the quantum mechanics of systems with a finite number of degrees of freedom to field theory in the framework of deformation quantization. We then relate our results to the analogous formulae in the operator formalism. We find that star products provide a natural setting for various forms of Wick's theorem, as well as the perturbative expansion of the time-evolution operator, which yields the Feynman rules of quantum field theory. In contrast to systems with a finite number of degrees of freedom, in field theory requirements of freedom from divergences and causality dictate the use of specific star products and quantization schemes.

\section{Star products in quantum mechanics}

In deformation quantization one starts from a classical system, that is, a Poisson manifold
$(M,\alpha)$, where $M$ is a smooth manifold and $\alpha$ the bivector which specifies the Poisson structure. The classical observables are smooth real-valued functions on $M$ which form, by pointwise multiplication, the commutative algebra of observables. Quantization involves a deformation of this algebra to a noncommutative associative algebra $\calA$ involving smooth complex-valued fuctions on $M$ which are multiplied by the use of a {\em star product}, defined as follows.

Let $\calA[[\hbar]]$ be the space of formal power series in $\hbar$ with coefficients in $\calA$. A star product $*$ is a bilinear map $\calA\times\calA\rightarrow \calA[[\hbar]]$ defined for $f,g\in\calA$
by
\begin{equation}
f*g=\sum_{n\ge 0}\left(  \frac{i\hbar}{2}\right)C_n(f,g),
\end{equation}
where the $C_n$ are bidifferential operators null on constants, such that:\newline
\noindent(i) $\sum_{r+s=t}C_r(C_s(f,g),h)=\sum_{r+s=t}C_r(f,C_s(g,h))$,\hspace*{\fill}\linebreak
\noindent(ii) $C_0(f,g)=fg$,\hspace*{\fill}\linebreak
\noindent(iii) $C_1(f,g)-C_1(g,f)=2\{ f,g \}.$\hspace*{\fill}\linebreak
By linearity $*$ has a unique extension to $\calA[[\hbar]]\times\calA[[\hbar]]\rightarrow\calA[[\hbar]]$. Property (i) expresses the associativity of the star product. Property (ii) is the classical limit: 
\begin{equation}
\lim_{\hbar\rightarrow 0}(f*g)(x)=f(x)g(x),
\end{equation}
with $x\in M$. 
Property (iii) expresses the correspondence principle:
\begin{equation}
\lim_{\hbar\rightarrow 0}\frac{1}{i\hbar}[f,g]_*=\{ f,g \},
\label{corrs}
\end{equation} 
with $[f,g]_*:=f*g-g*f$. In physical applications one is interested in {\em Hermitian} star products: $\overline{f*g}=\bar{g}*\bar{f}$, where $\bar{f}$ is the complex conjugate of $f$.

Two star products $*$ and $*'$ are said to be {\em c-equivalent} if there exists an invertible  {\em transition operator} $T$: a formal power series
\begin{equation}
T=\sum_{n\ge 0}(i\hbar)^nT_n,
\end{equation}
where $T_0=id$, and for $n>0$ the $T_n$ are bidifferential operators which are null on constants, such that
\begin{equation}
f*'g=T^{-1}((Tf)*(Tg)).
\end{equation}
Different choices of the star product correspond to different quantization schemes.

We shall be concerned in this paper with the case $M=\msr^{2m}$. In this case the coefficients of the Poisson structure $\alpha^{ij}$
may be taken as constants, and a star product, the {\em Moyal product} of the functions $f$ and $g$, is given by
\begin{equation}
(f\starM g)(x)=e^{M_{12}} f(x_1)g(x_2)
|_{x_1=x_2=x},
\label{Moyal}
\end{equation}
where 
\begin{equation}
M_{12}=\left( \frac{i\hbar}{2} \right) \sum_{i,j=1}^{2m}\alpha^{ij}
\frac{\partial}{\partial x_1^i}
\frac{\partial}{\partial x_2^j}\ ,
\end{equation}
and $x^i_\alpha$ $(i=1,\ldots,2m)$ is the $i$-th component of phase space point $x_\alpha$.
In canonical coordinates, $x=(q_1,\ldots,q_m,p_1,\ldots,p_m)$, 
$M_{12}$ is proportional to the Poisson bracket operator:
\begin{equation}
M_{12}=\left( \frac{i\hbar}{2} \right)
\sum_{i=1}^m \left(
\frac{\partial}{\partial q_1^i}
\frac{\partial}{\partial p_2^i}
-
\frac{\partial}{\partial p_1^i}
\frac{\partial}{\partial q_2^i}\right).
\end{equation}

For the rest of this section we will restrict our considerations for notational convenience to one-particle systems,
i.e. $m=1$ in $\msr^{2m}$. Furthermore, we use {\em holomorphic coordinates}:
\begin{equation}
a=\sqrt{\frac{m\omega}{2}}\left( q+\frac{ip}{m\omega}\right),\ \ 
 \bar{a}=\sqrt{\frac{m\omega}{2}}\left( q-\frac{ip}{m\omega}\right),
\label{hol}
\end{equation}
where $\omega$ is a constant which we shall later interpret as an oscillator frequency, and $m$ is a mass parameter. 
In these coordinates the Moyal product is
\begin{equation}
(f\starM g)(a,\bar{a})=e^{M_{12}}f(a_1,\bar{a}_1)g(a_2,\bar{a}_2)
\Big|_{
\begin{array}{l}
{\scriptscriptstyle a_1=a_2=a}\\
{\scriptscriptstyle \bar{a}_1=\bar{a}_2=\bar{a}}
\end{array}
},
\label{Moyal2}
\end{equation}
with
\begin{equation}
M_{12}=\frac{\hbar}{2}(\partial_{a_1}\partial_{\bar{a}_2}-\partial_{\bar{a}_1}\partial_{a_2}),
\label{Mcontraction}
\end{equation}
and $\partial_a=\partial /\partial a,\ \ \partial_{\abar}=\partial/\partial \abar$ .
For $M=\msr^{2m}$ all star products are c-equivalent to the Moyal product. For example, the {\em normal product} 
\begin{equation}
(f\starN g)(a,\bar{a})=e^{N_{12}}f(a_1,\abar_1)g(a_2,\abar_2)\Big|_{\begin{array}{l}{\scriptscriptstyle a_1=a_2=a}\\
{\scriptscriptstyle \bar{a}_1=\bar{a}_2=\bar{a}}\end{array}}\ ,
\end{equation}
with $N_{12}=\hbar \partial_{a_1}\partial_{\abar_2}$, 
is related to the Moyal product by the transition operator
\begin{equation}
T=e^{-\frac{\hbar}{2} \partial_a\partial_{\bar{a}}}.
\end{equation}

The Moyal product of $r$ functions is
\begin{equation}
f_1\starM f_2\starM \cdots \starM f_r= \exp\left( \sum_{i<j}M_{ij} \right) \prod_{m=1}^r f_m(a_m,\abar_m)
\Big|_{
\begin{array}{l}
{\scriptscriptstyle a_m=a}\\{\scriptscriptstyle \bar{a}_m=\bar{a}}
\end{array}
}.
\label{W0}
\end{equation}
There is a similar formula for the normal product. For functions $f_m$ which are {\em linear} in
$a$ and $\abar$,
\begin{equation}
f_m(a,\abar)=A_m a + B_m \abar,
\label{fm}
\end{equation}
the star product may be written in the form of a {\em Wick theorem}.
For example, the star product of four such functions can be written 
by expanding the exponential: 
\begin{eqnarray}
f_1\starM f_2\starM f_3\starM f_4 & = & f_1 f_2 f_3 f_4
+G_{12}\ (f_3 f_4)+G_{13}\ (f_2 f_4)+G_{14}\ (f_3 f_3) \nonumber\\
& &+G_{23}\ (f_1 f_4)+G_{24}\ (f_1 f_3)
+G_{34}\ (f_1 f_2)\nonumber\\
& &+G_{12}G_{34}+G_{13}G_{24}+G_{14}G_{23},
\end{eqnarray}
where the {\em contractions}
\begin{equation}
G_{ij}= M_{ij}f_i  f_j=\frac{\hbar}{2}(A_iB_j-A_jB_i)
\label{contractions}
\end{equation} 
are constants.
We then have the relation 
\begin{equation}
M_{ij}=G_{ij}\frac{\partial}{\partial f_i}\frac{\partial}{\partial f_j},
\label{contract}
\end{equation}
and Eq. (\ref{W0}) may be written as
\begin{equation}
f_1\starM f_2\starM \cdots \starM f_r= \exp\left( 
\sum\limits_{i<j}G_{ij}\frac{\partial}{\partial f_i}\frac{\partial}{\partial f_j} 
\right) \prod_{m=1}^r f_m\ .
\label{W4}
\end{equation}
It should be clear from the above that not only the original form \cite{Wick}, but also the various
{\em generalized} Wick theorems which have been discussed in the literature \cite{AW,HLS} are direct consequences of the structure of the relevant star products. 

\section{The forced harmonic oscillator}

We first give a brief review of the deformation quantization treatment of
 the simple one-dimensional free harmonic oscillator. The Hamilton function for this system is
\begin{equation}
H=\omega a \abar.
\label{oscil}
\end{equation}
The time development of a quantum system is governed by the {\em time evolution function},
for a system with a time-independent Hamilton function this is given by the {\em star exponential}: 
\begin{equation}
U(t,0)={\rm Exp}_* (Ht)=\sum_{n=0}^\infty \frac{1}{n!}\left( \frac{-i t}{\hbar}\right)^n (H*)^n ,
\end{equation}
where $(H*)^n=\underbrace{H*H*\cdots *H}_{{\it n}\ {\rm times}}$. 
We choose the quantization scheme characterized by the normal product. The star exponential then satisfies the differential equation
\begin{equation}
i\hbar \frac{d}{dt}
U(t,0)=H\starN U(t,0)=( H+\hbar\omega \abar \partial_{\abar} )\ U(t,0).
\end{equation}
The solution is
\begin{equation}
U(t,0)=e^{ -a\abar/\hbar }
\exp\left[
e^{-i\omega t} a\abar/\hbar \right].
\label{Sol}
\end{equation}
Developing the last exponential leads to the Fourier-Dirichlet expansion
\begin{equation}
U(t,0)=\sum\limits_{n=0}^\infty \pi_n^{(N)}\ e^{-i E_n t/\hbar},
\label{FD}
\end{equation}
where $\pi_n^{(N)}$ is the {\em projector} associated with the energy $E_n$. For the free harmonic oscillator
we read off from Eqs. (\ref{Sol}) and (\ref{FD})
\begin{eqnarray}
\pi_0^{(N)} & = &  e^{-  a \bar{a}/\hbar },\\
\pi_n^{(N)} & = & \frac{1}{\hbar ^n n!} \ \pi_0^{(N)}\ \bar{a} ^n a ^n, 
\label{normal}\\
E_n  & = &  n\hbar \omega,
\label{false}
\end{eqnarray}
where the projectors are normalized according to
\begin{equation}
\frac{1}{2\pi\hbar}\int d^2a \ \pi_n^{(N)}=1,
\end{equation}
and $d^2 a=da\,d\abar$. Note the absence of a zero-point energy in the spectrum. 
If we use the Moyal quantization scheme we obtain the spectrum
\begin{equation}
E_n= \left(n+\scriptstyle{\frac{1}{2}}\right)\hbar\omega,
\label{zero}
\end{equation}
including the usual zero-point energy \cite{HH}.

Now consider a harmonic oscillator in interaction with a time-dependent external source $J(t)$. 
The Hamilton function is
\begin{equation}
H=\omega a \abar -J(t)\abar - \bar{J}(t)a.
\end{equation}
The time evolution function $U_J$ is characterized by the differential equation
\begin{equation}
i\hbar \frac{d}{dt} U_J (t,t_i)=[ H + \hbar (\omega\abar-\bar{J}(t))\partial_{\abar}]
\ U_J(t,t_i).
\end{equation}
The solution is
\begin{eqnarray}
U_J(t_f,t_i) & = & e^{-a\abar/\hbar}
\exp\bigg[ \frac{1}{\hbar}
a\abar e^{i\omega(t_f-t_i)}
      +\frac{i}{\hbar} a  e^{i\omega t_f} \int_{t_i}^{t_f}ds e^{-i\omega s}\bar{J}(s)\nonumber\\
&{}&  +\frac{i}{\hbar}\abar e^{-i\omega t_f} \int_{t_i}^{t_f}ds e^{i\omega s}J(s) 
       -\frac{1}{\hbar}\int_{t_i}^{t_f}ds\int_s^{t_f} du\ e^{i\omega(u-s)} J(s)\bar{J}(u)  
\bigg].
\label{Source}
\end{eqnarray}

In the scattering situation we require that the source term become negligible as $|t|\rightarrow \infty$. The asymptotic dynamics is then governed by the Hamilton function for the free system: $H_0=H|_{J=0}$. The {\em scattering function} relates the asymptotic in-states to
the asymptotic out-states, where the source term is effective only in an at first limited time 
interval
 $-T<t<T$:
\begin{equation}
S[J]=\lim_{T\rightarrow\infty}U(0,T)\starN U_J(T,-T)\starN U(-T,0).
\end{equation} 
The phase space variables $a$ and $\abar$ develop in time under the influence of the free time 
evolution function as solutions of the free equations of motion, and for a general function 
$f(a,\abar)$ one finds
\begin{equation}
U(0,T)\starN f(a,\abar)\starN U(-T,0)=f\left( ae^{-i\omega T},\abar e^{i\omega T}\right).
\end{equation}
For the harmonic oscillator with a time-dependent source this yields, from Eq. (\ref{Source}):
\begin{equation}
S[J]=\exp{
\left[
    \frac{i}{\hbar}a\bar{j}(\omega)
+\frac{i}{\hbar}\abar j(\omega)
-\frac{1}{2\hbar}\int\limits_{-\infty}^\infty \int\limits_{-\infty}^\infty ds\, du\, e^{-i\omega |s-u|} J(s)\bar{J}(u)
\right]
},
\label{gen}
\end{equation}
where $j(\omega)=\int ds J(s)e^{i\omega s}$ is the Fourier transform of $J(s)$. Let
$\phi(t)=ae^{-i\omega t}+\abar e^{i\omega t}$,
and let $J(t)$ be real. Then Eq. (\ref{gen}) may be written as
\begin{equation}
S[J]=e^{ \frac{i}{\hbar}\int dt J(t)\phi(t) }
\exp\left[ -\frac{1}{2\hbar^2}\iint dt dt' J(t)D_F(t-t')J(t')\right], 
\label{SJ1}
\end{equation}
with
\begin{equation}
D_F(t)=\hbar\left[ \theta (t)e^{-i\omega t}+\theta (-t)e^{i\omega t} \right]\ .
\label{F0}
\end{equation}
We shall see below that the scattering function $S[J]$ corresponds to the scattering operator of quantum field theory, and $D_F(t)$ corresponds to the Feynman propagator (this correpondence is the reason for the factor $\hbar$ in the above equation). The {\em generating functional} is the 
vacuum expectation value of the scattering operator; in the phase space formalism this quantity 
is calculated as
\begin{equation}
Z_0[J]=\frac{1}{2\pi\hbar}\int d^2a\,S[J]\starN \pi_0^{(N)}=\exp\left[-\frac{1}{2\hbar^2}\iint dt dt' J(t)D_F(t-t')J(t')\right].
\label{genfunc}
\end{equation}
We may also calculate off-diagonal matrix elements of the scattering operator by making use of the {\em Wigner functions}
\begin{equation}
\pi_{m,n}^{(N)}=\frac{1}{\sqrt{\hbar^m\hbar^n m!n! }}\pi_0^{(N)}\abar^m a^n.
\end{equation}
Obviously, the Wigner functions are straightforward generalizations of the projectors: 
$\pi_n^{(N)}=\pi_{n,n}^{(N)}$. The transition amplitude for the system to go from the ground state to the state with energy $E_n$ under the influence of the source is
\begin{equation}
{\rm Amp}(0\rightarrow n)=\frac{1}{2\pi\hbar}\int d^2a\,\pi_{0,n}^{(N)}\starN S[J] \starN \pi_0^{(N)}=\frac{(i j(\omega))^n}{\hbar^{n/2}\sqrt{n!}}e^{-|j(\omega)|^2/2\hbar}.
\label{Amp}
\end{equation}
In this equation we have used the decomposition of the propagator into its real and imaginary parts,
\begin{equation}
D_F(t)=\hbar\left[ \cos{\omega t}-i\epsilon(t)\sin{\omega t} \right]\ ,
\end{equation}
to write
\begin{multline}
\exp\left[ -\frac{1}{2\hbar^2}\iint dt dt' J(t)D_F(t-t')J(t') \right]\\
=e^{-|j(\omega)|^2/2\hbar}
\exp\left[ -\frac{1}{2\hbar^2}\iint dt dt' J(t)D_R(t-t')J(t') \right],
\label{eq:E}
\end{multline}
where
\begin{equation}
D_R(t)=\hbar\theta(t)(e^{-i\omega t}-e^{i\omega t})=-2i\hbar\theta(t)\sin(\omega t)
\end{equation}
is the {\em retarded propagator}, and then discarded the unphysical phase represented by the second factor.
The probability for the above transition is
\begin{equation}
P_n=|{\rm Amp}(0\rightarrow n)|^2=\frac{|j(\omega)|^{2n}}{\hbar^n n!}e^{-|j(\omega)|^2/\hbar}.
\end{equation}
This corresponds to the well-known Poisson distribution for the number of emitted quanta in the field theoretical context:
\begin{equation}
P_n=e^{-\bar{n}}\frac{\bar{n}^n}{n!}\ ,
\end{equation}
where $\bar{n}$ is the average number of emitted quanta: 
\begin{equation}
\bar{n}=\sum\limits_{n=0}^{\infty}n P_n=|j(\omega)|^2/\hbar.
\end{equation} 

Our aim in this section was to demonstrate that one can calculate quantities of physical interest 
working exclusively at the level of phase space, that is, within the framework of deformation quantization. These quantum mechanical results may be taken over to the field theoretical context by a
formal extension to the case where the system considered has an infinite number of degrees of freedom. This will be shown in Sec. 5. The reader will probably recognize the results more easily when they are written in the conventional notation of the operator formalism. For this reason we discuss in the following section how our formulae can be transcribed into the operator language. 
 
\section{The operator formalism}

In the operator formalism we associate a linear operator in Hilbert space to each phase space function, in a manner which depends on the chosen quantization scheme. Note that our convention for the holomorphic variables $a,\abar$ in Eq. (\ref{hol}) implies $[a,\abar]_*=\hbar$ and thus $[\hat{a},\hat{a}^\dagger]=\hbar$, in contrast to the usual Fock space normalization. 
When we use the Moyal product scheme we obtain the associated {\em Weyl-ordered} operator (symmetrically ordered in $\hat{a}$ and $\hat{a}^\dagger$). For the normal product scheme we obtain the {\em normal ordered} operator (all $\hat{a}^\dagger$'s to the left of all $\hat{a}$'s). We denote the mapping from a given phase space function to its associated operator by $\hat{f}(\hat{a},\hat{a}^\dagger)=\Theta_X(f(a,\abar))$, where $X$ denotes the quantization scheme: $X=M$ for the Moyal product scheme, $X=N$ for the normal product scheme. We shall call this mapping the {\em Weyl transform}.

A product of operators is the Weyl transform of the star product of the corresponding phase 
space functions \cite{Groenewold}. For example, for the Moyal product scheme:
\begin{equation}
\hat{f}_1\cdots \hat{f}_r=\Theta_M\{(f_1\starM\cdots \starM f_r)(a,\abar)\}= \Theta_M\left\{ 
\exp\left(\sum_{i<j}M_{ij}\right)
\prod_{m=1}^{r} f_m(a_m,\abar_m)
\Big|_{\begin{array}{l}{\scriptscriptstyle a_m=a}\\
{\scriptscriptstyle \bar{a}_m=\bar{a}}\end{array}}
 \right\}.
\end{equation}
For a quantization scheme which is c-equivalent to the Moyal scheme we use the corresponding contraction factors $X_{ij}$ instead of the Moyal contraction factors
$M_{ij}$. We may write $X_{ij}=X_{\{ij\}}+M_{ij}$, where $X_{\{ij\}}=\frac{1}{2}(X_{ij}+X_{ji})$ is the symmetric part of $X_{ij}$,
 since the antisymmetric part is fixed for all c-equivalent star products by property (iii) of the definition.

The {\em time-ordered product} of $r$ time-dependent operators is given by the prescription
\begin{equation}
{\cal T}\{
\hat{f}_1(t_1)\cdots \hat{f}_r(t_r)
\} 
= \Theta_X \left\{ \exp\left(
\sum_{i<j} (X_{\{ij\}}+\epsilon(t_i-t_j)M_{ij})
\right)
\prod_{m=1}^r f_m(a_m,\abar_m,t_m)
\Big|_{\begin{array}{l}{\scriptscriptstyle a_m=a}\\
{\scriptscriptstyle \bar{a}_m=\bar{a}}\end{array}}
\right\},
\label{W1}
\end{equation}
since the transposition of two operators leaves $X_{\{ij\}}$ invariant, while the signs
of $\epsilon(t_i-t_j)$ and of $M_{ij}$ reverse.
For the case of normal ordering we may write the exponent in Eq. (\ref{W1}) as
\begin{align}
T_{ij}=N_{\{ij\}}+\epsilon(t_i-t_j)M_{ij}&= \frac{\hbar}{2}\left[
( \partial_{a_i}\partial_{\abar_j}
 +\partial_{\abar_i}\partial_{a_j} )
+\epsilon(t_i-t_j) 
( \partial_{a_i}\partial_{\abar_j} 
 -\partial_{\abar_i}\partial_{a_j} )\right] \notag\\
&=
 \frac{\hbar}{2}\left[ (1+\epsilon(t_i-t_j))\partial_{a_i}\partial_{\abar_j}
+(1-\epsilon(t_i-t_j))\partial_{\abar_i}\partial_{a_j}\right]\notag\\
&=
 \hbar\left[ \theta(t_i-t_j)\partial_{a_i}\partial_{\abar_j}
        +\theta(t_j-t_i)\partial_{\abar_i}\partial_{a_j}\right] .
\label{X}
\end{align}

Suppose now that the functions $f_m$ are linear in $a_m$ and $\abar_m$, and have a periodic time dependence:
\begin{equation}
f_m(t)=A_m a_m e^{-i\omega t}+B_m \abar_m e^{i\omega t}.
\end{equation}
By Eq. (\ref{contractions}) the relevant contractions are
\begin{equation}
D_{ij}( t-t' ) = T_{ij} f_i(t) f_j(t')=
 \hbar \left[ A_i B_j \theta(t-t')e^{-i\omega (t-t')}
+ A_j B_i \theta(t'-t)e^{i\omega (t-t')}\right],
\label{F1}
\end{equation}
which is a generalization of the expression in Eq. (\ref{F0}). We write, in analogy to 
Eq. (\ref{contract}),
\begin{equation}
T_{ij}=\iint dt\,dt'\,\frac{\delta}{\delta f_i(t)} D_{ij}(t-t') 
             \frac{\delta}{\delta f_j(t')}\ ,
\end{equation}
where the $\delta/\delta f(t)$ are functional derivatives.
For the operators
\begin{equation}
\hat{f}_m(t)=A_m \hat{a} e^{-i\omega t}+B_m \hat{a}^\dagger e^{i\omega t}
\end{equation}
we get a quantum mechanical form of Wick's theorem by inserting these expressions into 
Eq. (\ref{W1}):
\begin{multline}
{\cal T} \{ \hat{f}_1(t_1)\cdots \hat{f}_r(t_r) \} \\
=\Theta_N \left\{
\exp
\left(
 \sum\limits_{i<j}^{}
\iint\, dt\, dt'\frac{\delta}{ \delta f_i(t) } D_{ij}(t-t')
\frac{\delta}{ \delta f_j(t') }
\right) 
 f_1(t_1)\cdots f_r(t_r) 
\Big|_{
\begin{array}{l}
{\scriptscriptstyle a_m=a}\\
{\scriptscriptstyle \bar{a}_m=\bar{a}}
\end{array}
}
\right\}.
\label{Wick}
\end{multline}
Since we have modified the star product contractions in Eq. (\ref{W1}) by the insertion of the $\epsilon(t_i-t_j)$ factor, the time-ordered product is {\em not} the Weyl transform of a star product. This can be seen from the fact that the time-ordered product is {\em symmetric} in its arguments, whereas the star products have an antisymmetric part fixed by property (iii) of their definition.

\section{Quantum field theory}

The generalization of the foregoing results to the field theoretical context is formally straightforward:
the free real scalar field is equivalent to an infinite collection of independent harmonic oscillators. 
We at first consider the system to be confined to a box of finite volume $V$.  
The representation of the free field in terms of the variables $a({\bf k}),\abar({\bf k})$ is
\begin{equation}
\phi(x)=\frac{1}{\sqrt{V}}\sum\limits_{{\bf k}} \frac{1}{\sqrt{2\omega_{\bf k}}}\left[
a({\bf k})e^{-ikx}+\abar({\bf k})e^{ikx} \right],
\label{cfield}
\end{equation}
where $\hbar k^0=\hbar \omega_{{\bf k}}=\sqrt{\hbar^2|{\bf k}|^2+m^2}$ is the energy of a single quantum of the field. The normalization of the field is fixed by the equal-time commutator
\begin{equation}
[\phi(x),\dot{\phi}(y)]_* \Big|_{x^0=y^0}=i\hbar \delta^{(3)}({\bf x}-{\bf y}).
\end{equation}
The Hamilton function is the generalization of Eq. (\ref{oscil}):
\begin{equation}
H=\sum\limits_{\bf k}\omega_{\bf k} a({\bf k})\abar({\bf k}). 
\end{equation} 
The vacuum state in the normal product scheme is
\begin{equation}
\pi_0^{(N)}=\prod_{{\bf k}}\left(e^{-a({\bf k})\abar({\bf k})/\hbar}\right)=
e^{-\sum\limits_{{\bf k}}a({\bf k})\abar({\bf k})/\hbar},
\end{equation}
and is normalized in the usual manner:
\begin{equation}
\int \prod_{\bf k}\left( \frac{d^2a_{{\bf k}}}{2\pi\hbar}\right)\pi_0^{(N)}=1.
\end{equation}
The vacuum expectation value of $H$ vanishes:
\begin{equation}
\int \prod_{\bf k}\left( \frac{d^2a_{\bf k}}{2\pi\hbar}\right) H\starN \pi_0^{(N)} =0. 
\end{equation}
Had we used the Moyal product quantization scheme we would have found an infinite vacuum energy, arising from the zero-point energy in the spectrum in Eq. (\ref{zero}). This fact has been used to argue that the normal product is the only admissible star product in the context of free field theory \cite{D1}. From now on we shall go over to the continuum normalization  of the fields:
\begin{equation}
\phi_m(x)=\int \frac{d^3k}{(2\pi)^{\frac{3}{2}}} \frac{1}{\sqrt{2\omega_{\bf k}}}\left[
a_m({\bf k})e^{-ikx}+\abar_m({\bf k})e^{ikx} \right]\ .
\label{rfield}
\end{equation}

To form the Moyal product of fields we first calculate the relevant contractions by generalizing
 Eq. (\ref{contractions}) to a system with an infinite number of degrees of freedom: 
\begin{eqnarray}
\frac{1}{2}D(x_1-x_2) & =&  \iiint d^3 k \frac{d^3 k_1 }{ (2\pi)^{\frac{3}{2}} }
                                         \frac{d^3 k_2 }{ (2\pi)^{\frac{3}{2}} }
\frac{1}{ \sqrt{ 2\omega_{{\bf k}_1}} }
\frac{1}{ \sqrt{ 2\omega_{{\bf k}_2}} } 
\frac{\hbar}{2}\left[ 
\frac{ \delta }{ \delta     a_1({\bf  k}) }
\frac{ \delta }{ \delta \abar_2({\bf  k}) }-
\frac{ \delta }{ \delta \abar_1({\bf k }) }
\frac{ \delta }{ \delta     a_2({\bf k }) }
\right] \notag\\
&&\ \ \ \times \left( a_1({\bf k}_1)e^{-ik_1 x_1}+\abar_1({\bf k}_1)e^{ ik_1 x_1 } 
\right)
\left( a_2({\bf k}_2)e^{-ik_2 x_2}+\abar_2({\bf k}_2)e^{ ik_2 x_2 } 
\right)
\notag\\
&=& \frac{1}{2}\left[ D^+(x_1-x_2)+D^-(x_1-x_2)\right]\ ,\notag\\
\label{Schwinger}
\end{eqnarray}
where
\begin{equation}
D^{\pm}(x)=\pm \int \frac{d^3 k}{(2\pi)^3}\frac{\hbar}{ 2\omega_{{\bf k}} }\ e^{\mp ikx}
\end{equation}
are the propagators for the components of positive and negative frequencies,
and $D(x)$ is the Schwinger function.
The Moyal product of the fields is then, in analogy to Eq. (\ref{W4}),
\begin{equation}
\phi(x_1)\starM \cdots \starM \phi(x_r)=
\exp\left[
\frac{1}{2}\sum\limits_{i<j}\iint d^4 x\,d^4 y\,\frac{\delta}{\delta \phi_i(x)}D(x-y)
                                \frac{\delta}{\delta \phi_j(y)}
\right]
\prod_{m=1}^r \phi_m(x_m)\Big|_{\phi_m=\phi}.
\label{Munendlich}                                  
\end{equation}
For the quantum field operators
\begin{equation}
\hat{\Phi}(x)=\int \frac{d^3 k}{(2\pi)^\frac{3}{2}}\frac{1}{\sqrt{2\omega_{\bf k}}}\left[
\hat{a}({\bf k})e^{-ikx}+\hat{a}^\dagger({\bf k})e^{ikx} \right]\ 
\end{equation}
we obtain
\begin{multline}
\hat{\Phi}(x_1)\cdots \hat{\Phi}(x_r)\\
=\Theta_M \left\{
\exp\left[
\frac{1}{2}\sum\limits_{i<j}\iint d^4 x\,d^4 y\,\frac{\delta}{\delta \phi_i(x)}D(x-y)
                                \frac{\delta}{\delta \phi_j(y)}
\right]
\prod_{m=1}^r \phi_m(x_m)\Big|_{\phi_m=\phi} \right\}\ .
\label{Mfields}                                  
\end{multline}

The time-ordered product of the quantum field operators is, by analogy to Eq. (\ref{Wick}),
\begin{multline}
{\cal T}\{ \hat{\Phi}(x_1)\cdots\hat{\Phi}(x_r) \}\\
=\Theta_N \left\{ \exp\left[
\sum\limits_{i<j}\iint d^4 x\,d^4 y\,\frac{\delta}{\delta \phi_i(x)}D_F(x-y)
\frac{\delta}{\delta \phi_j(y)}
\right]
\prod_{m=1}^r \phi_m(x_m)\Big|_{\phi_m=\phi}\right\}.     
\label{Wf}
\end{multline}
Here $D_F$, the {\em Feynman propagator}, is given by the infinite dimensional generalization of Eq. (\ref{F1}):
\begin{eqnarray}
D_F (x_1-x_2) & = & \iiint d^3 k \frac{d^3 k_1 }{(2\pi)^\frac{3}{2}} \frac{d^3 k_2}{(2\pi)^\frac{3}{2}} 
\frac{1}{ \sqrt{ 2\omega_{{\bf k}_1}} } \frac{1}{ \sqrt{ 2\omega_{{\bf k}_2}} } 
\nonumber\\
&& \times \hbar\left[ 
\theta(t_1-t_2)
\frac{ \delta }{ \delta     a_1({\bf  k}) }
\frac{ \delta }{ \delta \abar_2({\bf  k}) }+
\theta(t_2-t_1)
\frac{ \delta }{ \delta \abar_1({\bf k }) }
\frac{ \delta }{ \delta     a_2({\bf k }) }
\right]\nonumber\\
&& \times
\left( a_1({\bf k}_1)e^{-ik_1 x_1}+\abar_1({\bf k}_1)e^{ ik_1 x_1 } 
\right)
\left( a_2({\bf k}_2)e^{-ik_2 x_2}+\abar_2({\bf k}_2)e^{ ik_2 x_2} 
\right)
\nonumber\\
&=&\theta(t_1-t_2)D^+(x_1-x_2)-\theta(t_2-t_1)D^-(x_1-x_2)).
\label{Fprop}
\end{eqnarray}
We may simplify Eq. (\ref{Wf}) by using the symmetry of the Feynman propagator, 
$D_F(x_1-x_2)=D_F(x_2-x_1)$; it becomes:
\begin{equation}
{\cal T}\{ \hat{\Phi}(x_1)\cdots\hat{\Phi}(x_r) \}=\Theta_N \left\{
\exp\left[
\frac{1}{2}\iint d^4 x\,d^4 y\,\frac{\delta}{\delta \phi(x)}D_F(x-y)
\frac{\delta}{\delta \phi(y)}
\right]
\phi(x_1)\cdots\phi(x_r) \right\}.
\label{W6}                                
\end{equation}
Note that in this case it is no longer necessary to use different fields which are set equal only after the differentiation; because of the symmetry the correct combinatorics are guaranteed by the Leibnitz rule for differentiation. Eq. (\ref{W6}) is the field-theoretic version of Wick's theorem.

The propagator for positive frequencies $D^+(x)$ is c-equivalent to $\frac{1}{2}D(x)$ by use of the transition operator
\begin{equation}
T=\exp\left[-\frac{1}{2}\iint d^4x\,d^4y\,\frac{\delta}{\delta \phi(x)}\frac{1}{2}\left[D^+(x-y)-D^- (x-y)\right]
\frac{\delta}{\delta \phi(y)}\right].
\end{equation} 
The time-ordered product for the field operators, Eq. (\ref{Wf}), is the Weyl transform of the expression which results from the Moyal star product, Eq. (\ref{Munendlich}), by replacing $\frac{1}{2}D$ by $D^+$, restricting the integration to positive times $x^0>y^0$, and symmetrizing.

For $n=2$ Wick's theorem is
\begin{equation}
{\cal T} \{ \hat{\Phi}(x_1)\hat{\Phi}(x_2) \}= \Theta_N \{ \phi(x_1)\phi(x_2) \}
+D_F(x_1-x_2).
\end{equation}
Since the vacuum expectation value of the normal product vanishes, this yields the familiar relation
\begin{equation}
D_F(x_1-x_2)=\langle0|{\cal T}\{\hat{\Phi}(x_1)\hat{\Phi}(x_2)\}|0\rangle.
\end{equation}

Wick's theorem may also be written in the form of a generating function:
\begin{equation}
{\cal T} \{ e^{ \frac{i}{\hbar}\int d^4 x  J(x)\hat{\Phi}(x) }   \}
=\Theta_N \{ e^{\frac{i}{\hbar}\int d^4 x  J(x)\phi(x) }   \}
\exp\left[ -\frac{1}{2\hbar^2}\iint d^4 x\,d^4 y J(x)D_F(x-y)J(y) \right],
\label{pert}
\end{equation}
where $J(x)$ is an external source. Eq. (\ref{Wf}) then results by expanding both sides of Eq. (\ref{pert}) in powers of $J$ and comparing coefficients. 
Note that
\begin{equation}
\hat{S}[J]={\cal T}\{e^{\frac{i}{\hbar}\int d^4 x  J(x)\hat{\Phi}(x) }   \}={\cal T}\{e^{ -\frac{i}{\hbar}\int d^4 x 
\hat{H}_{\rm int}(x)  }   \}
\label{scattop}
\end{equation}
is the {\em scattering operator} of quantum field theory, so that Eq. ({\ref{pert}}) is the perturbation expansion of the scattering operator for this interaction. This is just the operator form of our previous result, Eq. (\ref{SJ1}), which was derived completely within the phase space formalism of deformation quantization theory. The generating functional for the perturbation series is, by Eq. (\ref{pert}),
\begin{equation}
Z_0[J]=\langle 0|\hat{S}[J]|0\rangle =\exp\left[ -\frac{1}{2\hbar^2}\iint d^4 x\,d^4 y J(x)D_F(x-y)J(y) \right],
\label{func}
\end{equation} 
in agreement with Eq. (\ref{genfunc}). When a self-interaction term is included in the interaction Hamiltonian, $\hat{H}_{\rm int}=-J\phi+V(\phi)$, the generating functional for the interacting theory becomes
\begin{equation}
Z[J]=\frac{1}{N}\ e^{ -\frac{i}{\hbar} \int d^4x\,V\left(\frac{\hbar}{i}\frac{\delta}{\delta J(x)}\right) }\ Z_0[J]\ ,
\end{equation}
where the normalization constant is $N=Z[J=0]$.

\section{Conclusions}

The generating functional for the free field theory with an arbitrary source is the starting point for the derivation of the Feynman rules of the interacting theory in any functional scheme, whether it is calculated by the use of path integrals, as in Feynman and Hibbs \cite{FH}, or by Schwinger's source theory \cite{Schwinger}. In this paper we have demonstrated that it is also possible to
calculate the generating functional by using exclusively the methods of deformation quantization. Hence we see that the framework of deformation quantization can encompass the whole range of quantum theory, from simple non-relativistic systems with a finite number of degrees of freedom to interacting relativistic quantum fields.

 It is remarkable that when one goes beyond the formal level, and asks whether the star products and products of distributions which occur in these calculations are well-defined, one is lead to the same choices of quantization schemes and propagators which here appear on purely formal grounds: The normal star product is the only well-defined product which can be used in the free field context \cite{D1}, and the positive frequency propagator $D^+$, which is c-equivalent to the Schwinger function, must be used in relativistic perturbation theory in order to avoid ill-defined products of distributions \cite{Fred}, and to eventually obtain the Feynman propagator which occurs in the time-ordered products. Physically, the necessity for the use of the Feynman propagator is related to the principle of relativistic causality \cite{Wein}.

\end{document}